\documentclass{aa} 
\usepackage{graphicx}
\begin{document} 
 
\title{The IC\,2118 association:
new T~Tauri stars in high-latitude molecular clouds} 
 
\author{M. Kun\inst{1} \and T. Prusti\inst{2} \and S. Nikoli\'c\inst{3,4} \and
 L. E. B. Johansson\inst{4} \and N. A. Walton\inst{5}} 
\institute{Konkoly Observatory, H-1525 Budapest, P.O. Box 67, Hungary 
\and  Astrophysics Missions Division, 
Research and Scientific Support Department of ESA, 
Postbus 299, NL-2200 AG Noordwijk, The Netherlands
\and Astronomical Observatory, Volgina 7, 11160 Belgrade 74, Serbia, Serbia and Montenegro
\and Onsala Space Observatory,  S-439 92 Onsala, Sweden
\and Institute of Astronomy, University of Cambridge, Madingley Road,
Cambridge CB3 0HA, UK}
\offprints{M. Kun,
\email{kun@konkoly.hu}} 
\date{Received  / Accepted }
  
\abstract{
We identified new pre-main sequence stars in the region of high-latitude
molecular clouds associated with the reflection nebula IC\,2118,
around $l \sim 208\degr$ and $b \sim -27\degr$.
The stars were selected as T~Tauri candidates in objective prism
plates obtained with the Schmidt telescope of Konkoly
Observatory. Results of spectroscopic follow-up observations,
carried out with the FLAIR spectrograph installed on the
UK Schmidt and with ALFOSC on Nordic Optical Telescope, are 
presented in this paper. Based on spectral types, presence of emission 
lines and lithium absorption line, we identified  five classical T~Tauri
stars and a candidate weak-line T~Tauri star projected on the molecular 
clouds, as well as two candidate pre-main sequence stars outside the
nebulous region. Using the near infrared magnitudes obtained from the 2MASS 
All Sky Catalog (IPAC~2003) we determined  the masses and ages of these
stars. We found that the five classical T~Tauri stars projected on the clouds
are physically related to them, whereas the other stars are probably background
objects. Adopting a distance of 210\,pc  for IC\,2118 (Kun et al.~2001) 
and using Palla \& Stahler's~(1999) evolutionary tracks we derived 
an average age of $2.5\times10^{6}$ yrs and a mass interval of 
0.4--1.0\,M$_{\sun}$ for the members of the IC\,2118 association.
}
\titlerunning{The IC\,2118 association}
\authorrunning{Kun et al.}
\maketitle

\section{Introduction}
\label{Sect_1}

Small molecular clouds at high galactic latitudes 
(Magnani, Blitz \& Mundy~\cite{MBM})  are usually
devoid of star formation. A well-studied exception is MBM~12, 
containing the young association of low-mass pre-main sequence (PMS) stars
MBM\,12A (e.g. Luhman~\cite{Luhman}). 
A less studied example of star forming high latitude molecular
cloud is MBM~21 which harbours an infrared source, IRAS~04591$-$0856,
associated with a faint nebulosity HHL\,17 (Gyulbudaghian, Rodr\'{\i}guez,
\& Mendoza-Torres~\cite{GRM}). Persi et al.~(\cite{Persi}) have shown that
HHL\,17 is a low-mass YSO  between the protostellar
and pre-main sequence evolutionary stage.
MBM~21 and 22 are projected at an angular distance of some 10 degrees
from the Orion~A molecular cloud. They lie at the southernmost part of
an extended reflection nebula, IC\,2118 (Witch Head Nebula), illuminated 
by $\beta$~Orionis (Rigel). 
A $^{12}$CO survey performed with the 4-meter NANTEN radio telescope and 
covering the whole area of IC\,2118 (Kun et al.~\cite{KAY}, hereinafter Paper~I) 
resulted in the detection of six molecular clouds in the bright region, 
including MBM~21 (G\,208.4$-$28.3) and MBM~22 (G\,208.1$-$27.5). 
The most massive member of this small group of clouds, G\,206.4$-$26.0, 
is not included in the MBM catalogue, but was studied by Bally et 
al.~(\cite{Bally91}) and Yonekura et al.~(\cite{Yonekura}). 
Low-mass star formation in G\,206.4$-$26.0 is 
indicated by the small group of nebulous stars RNO\,37 
(Cohen~\cite{RNO}), with H$\alpha$ emission in the two 
brightest members (Nakano, Wiramihardja \& Kogure~\cite{NWK}).
The northern of these two stars
coincides with the IRAS source 05050$-$0614, having spectral energy 
distribution indicative of a PMS star (Paper~I).
Yonekura et al.'s~(\cite{Yonekura}) $^{12}$CO, $^{13}$CO and C$^{18}$O 
studies have shown that both clouds associated with IRAS point sources
contain high-density cores, suitable for forming low-mass 
stars.

IC\,2118 is a part of the Orion region surveyed for weak-line 
T~Tauri stars by Alcal\`a et al.~(\cite{A96}) on the 
basis of ROSAT all-sky survey. One wTTS of this sample, 
RXJ\,0502.4$-$0744 is projected on the reflection nebula, and another
one, RXJ\,0507.8$-$0931 is located at about 1.5 degrees to the east of it. 

The visual appearance of the clouds associated with IC\,2118
suggests their interaction with Orion OB\,1 (e.g. Ogura \& Sugitani~\cite{OS}), 
therefore they are usually thought to be as distant as the Orion~A
and B molecular clouds, i.e. $\sim$460\,pc. 
The radial velocities of the clouds, however, are more negative
($-5.30  < v_\mathrm{LSR} < +4.8$\,km\,s$^{-1}$) than both those of the main 
clouds Orion~A and Orion B, (3\,km\,s$^{-1} \lse v_\mathrm{LSR} \lse 11$\,km\,s$^{-1}$, 
Bally~(\cite{Bally89}); Aoyama et al.~(\cite{AMY})) and the 
bright stars of the Orion~OB1 association ($v_\mathrm{LSR} \sim$ +5\,km\,s$^{-1}$, 
Brown, de Geus \& de Zeeuw~(\cite{BGZ})). 
This velocity pattern  suggests that whereas the Orion~A and B molecular 
clouds are situated in the receding hemisphere of
the expanding interstellar structure around Orion~OB1 (Orion--Eridanus 
Bubble, Brown, Hartmann \& Burton~(\cite{BHB})), the 
IC\,2118 clouds belong to its approaching side. 
These small clouds therefore are probably closer to us than
the expansion centre of the Bubble, Ori~OB1a (336$\pm$16\,pc, de Zeeuw et 
al.~\cite{ZHB}). Considering the cometary shapes of the clouds, 
Bally et al.~(\cite{Bally91}) proposed that they are actually located inside the  
Bubble, whose radius is about 140\,pc (Brown et al.~\cite{BGZ}).
Based on literature data, Kun et al.~(\cite{KAY}) adopted 210\,$\pm20$\,pc 
for the most probable distance of IC\,2118. This result 
implies that the clouds are situated inside the Orion--Eridanus Bubble, 
and close to its surface nearest to us.

The age sequence of OB subgroups of Orion~OB\,1 
(e.g. Brown et al.~\cite{BGZ}) as well as star formation 
observed in some cometary globules (e.g. Stanke et al.~\cite{SSGS})
suggest that interactions of high-mass stars with the interstellar
medium have played a significant role in forming the present appearance 
of the region. Several observed properties of the
IC\,2118 region suggest that low mass star formation has been triggered 
here by the Orion--Eridanus Bubble. We performed a search for 
additional PMS stars in order to 
explore the star forming history of the region. 
Objective prism Schmidt plates were used to search
for H$\alpha$ emission stars, and then  spectroscopic follow-up
observations of the candidates were carried out in order to establish
their nature.

In this paper we present the results of our spectroscopic survey.
We describe our observations and data analysis in Section~\ref{Sect_2}. 
Results on the new PMS stars are shown in Sect.~\ref{Sect_3}.
A brief summary of the paper is given in Sect.~\ref{Sect_4}. 
As all of our target stars are included in the 2MASS All Sky 
Catalog~(IPAC~\cite{2MASS}), we use the 2MASS source designation 
for identifying our objects.

\section{Observations}
\label{Sect_2}

\subsection{Objective prism observations}
\label{Sect_2.1}

In order to find the possible classical T~Tauri stars 
in the region of IC\,2118 we performed an
objective prism Schmidt survey for  H$\alpha$ emission stars.
Observations were carried out in 1988/1989 
at Piszk\'estet\H o mountain station of Konkoly Observatory, 
with the 60/90/180\,cm Schmidt telescope equipped with a 5~degree 
objective prism. The field of view of the telescope was a circle
of five degrees in  diameter. The whole extent of the reflection
nebulosity was covered by two, partly overlapping, fields centred on
$\alpha(1950)=4^{\rm h}58^{\rm m}$,\/ $\delta(1950)=-8^{\rm o}20'$
and $\alpha(1950)=5^{\rm h}4^{\rm m}$,\/ $\delta(1950)=-6^{\rm o}30'$,
respectively. The exposures were taken on Kodak 098-02 and 103a-F
emulsions, through an RG1 filter in order to separate the spectral
region around the H$\alpha$ line. The exposure times were 60, 72 and
90 minutes, providing a limiting magnitude of about $R \sim 16$~mag
 according to our previous studies of this type (Kun~\cite{Kun1}, \cite{Kun2};
Kun \& P\'asztor~\cite{KP}). As the region is always 
seen at large zenith distances from Piszk\'estet\H o
($47\degr55\arcmin$ North), identification of the H$\alpha$ line was
rather uncertain due to the strong sky background. In order not to miss any 
relevant object, we selected for further study all the dubious cases,
63 stars in all. Equatorial coordinates of the selected objects 
were computed from their positions, measured on the 
objective prism plates,  using the H$\alpha$ line as a reference position
along the prism spectra. Their $R$  magnitudes were estimated from their diameters
in the red POSS prints. As the equatorial coordinates obtained from the 
objective prism spectra are uncertain to 1--3\arcsec\  (Kun~\cite{Kun1}), we used the list of 
suspected H$\alpha$ emission stars obtained in this manner together
with their finding charts as input data  for spectroscopic follow-up
observations.

\subsection{Spectroscopy}
\label{Sect_2.2}

Spectroscopic follow-up observations of the PMS star candidates 
selected from the objective prism observations were carried out 
at two different epochs and using 
two different instruments. A sample of 40 stars, brighter than 
about $R \sim 14.5$\,mag and distributed over a field of $6\degr \times 6\degr$ 
centred on RA(1950)=5$^{\mathrm h}0^{\mathrm m}$ and D(1950)=$-8\degr0\arcmin$, 
were observed using the FLAIR multi-fiber spectrograph 
installed on the UK Schmidt telescope on 15th December, 1993, 
at poor weather conditions. A pair of frames 
were taken through the low-dispersion grating G\,300B,
covering the spectral region between 3800 and 6600\,\AA, and another
pair using the high dispersion grating G\,1200R, covering the  
wavelength region 6000--6800\,\AA, each with an exposure time of 
3000\,sec. Domeflat and twilight frames were 
also obtained for calibration purposes. Neon and Cd-Hg lamp spectra 
were observed before and after the stellar frames for wavelength 
calibration. After the standard CCD reduction procedures, performed  
in {\sc IRAF}\footnote{IRAF is distributed by the National Optical Astronomy
Observatories, which is operated by the Association of Universities
for Research in Astronomy, Inc., under contract to the National
Science Foundation.}, the images taken through the same grating were coadded.
Individual spectra were extracted using the {\sc IRAF} task `dofibers'. 
The spectral resolution, estimated from the FWHM of the neon
lines, was $\lambda / \Delta \lambda \approx 500$ at $\lambda=5000$\,\AA\ 
for the G\,300\,B spectra, and $\lambda / \Delta \lambda \approx 11000$ at 
$\lambda=6560$\,\AA\  for the G\,1200\,R spectra.
Due to the small angular separation of the components of RNO\,37 only 
its brightest member could be observed during this observing run. 

Independently, the PMS star candidates projected on the illuminated 
clouds were observed using the ALFOSC spectrograph installed on the
Nordic Optical Telescope in La Palma in January 2000. The ALFOSC 
spectra were taken through grism~8, giving a dispersion of 1.5\,\AA/pixel
over the wavelength region 5800--8350\,\AA. Using a 1\arcsec\  slit
the spectral resolution was $\lambda / \Delta \lambda \approx 1000$ at 
$\lambda=6560$\,\AA.
Spectra of helium and neon lamps were observed before and 
after each stellar observation for wavelength calibration. We observed a 
series of spectroscopic standards for spectral classification purposes. 
We reduced the spectra using standard {\sc IRAF} routines.
The journal of the spectroscopic observations is given in Table~\ref{Tab1}.

{\small
\begin{table*}
\caption{Journal of spectroscopic observations}
\label{Tab1}
\begin{flushleft}
\begin{tabular}{ll@{\hskip2mm}l@{\hskip2mm}l@{\hskip2mm}c@{\hskip2mm}c@{\hskip2mm}l@{\hskip2mm}c}
\hline
\hline
\noalign{\smallskip}
Telescope & Date & Spectrograph & Grating/ &  Disp. & Sp. range & $\lambda/\Delta \lambda$
 & N  \\
  & & & Grism & (\AA/pix) & (\AA) \\
\noalign{\smallskip}  
\hline 
\noalign{\smallskip}
UK Schmidt &  15 Dec 1993 & FLAIR & 300B & 4.8 & 3800--6600 & ~~500 & 40 \\
UK Schmidt &  15 Dec 1993 & FLAIR & 1200R & 0.35 &  6000--6800 & 11000 & 40 \\
NOT        & 2--7 Jan 2000 & ALFOSC & Grism 8 & 1.5 & 5825--8350 & ~1000 & 10  \\
\noalign{\smallskip}
\hline
\end{tabular}
\end{flushleft}
\end{table*}}

\subsection{Spectral classification}
\label{Sect_2.3}

\subsubsection{FLAIR spectroscopy}
\label{Sect_FLAIR}

The wide wavelength range of FLAIR spectra taken through the grating 300\,B 
contains several features suitable for spectral classification. 
Actually the S/N of the blue part of the spectra (at wavelengths
shortward of H$\beta$) was too low due to the low sensitivity of
the CCD in this spectral region. Useful features for classifying 
late type stars in the 5000--6600\,\AA\ region are 
the MgI lines at 5164--5173, NaI 5890 and 5896, CaI at 6162 
and CaH at 6496\,\AA\ for G--K types, as well TiO bands at 5167, 5449, 
5862 and 6159\,\AA\  for the M-type stars.  Considering that  
our 300\,B spectra had nearly the same resolution as the  
spectrophotometric standards of Pickles'~(\cite{Pickles}) 
spectrum library, we utilized this data base in determining the 
spectral types of the observed stars. After converting 
the relevant parts of the published data files into 
{\sc IRAF\/} images and  normalizing both the observed 
and standard spectra to the continuum in the same manner, 
we calibrated several spectral features against the spectral type 
by measuring their equivalent widths on a grid of standard spectra.
The accuracy of the spectral classification is $\pm2$ subclasses.
Based on the presence of H$\alpha$ emission
and G, K or M spectral type seven candidate pre-main 
sequence stars were identified in the FLAIR G\,300\,B image over 
the whole field of view of the instrument, five of which were projected 
on IC\,2118. In addition to the above criteria  presence of forbidden 
emission lines as evidence of accretion, and/or strong LiI\,6707 
absorption line, an important indicator of youth (Bodenheimer~\cite{Bodenheimer})
are required for establishing the PMS nature of the stars.
It has to be noted as
well that the shape and equivalent width of the H$\alpha$ line 
in the fiber spectra are uncertain due to the fact that the same average 
sky spectrum was subtracted from each stellar spectrum, in spite of the
variable H$\alpha$ background throughout the field
of view. Therefore presence of the lithium line is a
primary criterion of the PMS nature of those target stars
which display only weak Balmer emission lines.
Spectra taken through the grating G\,1200\,R have, in principle, sufficiently
high resolution  to detect the lithium line and separate it from the neighbouring
CaI\,$\lambda$6718\,\AA\ line.  Most of our stars have been proved, however,
underexposed in the G\,1200\,R image, and thus the S/N of their 
spectra was insufficient for measuring reliably the LiI\,6707 equivalent 
width, a key indicator of age for G and K type PMS stars. 
Therefore we used the lithium line as a criterion 
so that we rejected as PMS objects the stars having H$\alpha$ 
emission and no lithium absorption, and classified stars with detected
LiI absorption and weak H$\alpha$ emission as {\em candidate\/} 
wTTS.

Spectra of the H$\alpha$ emission stars found during this observing run
are shown in Fig.~\ref{Fig1}. The left panels show the low resolution 
spectra over the wavelength region 4800--6600\,\AA\, and the right
panels show the wavelength interval 6540--6740\,\AA\ obtained with
the high resolution grating G 1200\,R. 
Table~\ref{Tab2} shows the results of the FLAIR observations.
Spectral types and H$\alpha$  equivalent widths are given, 
and we indicate the detection of other emission lines and 
LiI\,$\lambda$6707 absorption in the spectra. {\sl J\/} magnitudes, {\sl J$-$H\/}
and {\sl H$-$K$_s$\/} colour indices of the  stars from the 2MASS All Sky 
Catalog (IPAC~\cite{2MASS}) are also listed. 

According to different classification criteria (e.g. Mart\'{\i}n~\cite{Martin};
White \& Basri~\cite{WB}) for PMS stars, 
2MASS~J\,05020630$-$0850467 and   05073060$-$0610597 
are classical T~Tauri stars.  The third candidate cTTS is
2MASS J\,05112460$-$0818320, projected far from the 
IC\,2118 molecular clouds. We detected a surprisingly strong variation 
in the spectrum of this star. The low resolution 
spectrum shows an early M-type star with weak H$\alpha$ emission, 
whereas one can see very strong H$\alpha$ and other emission 
features such as [NI]~$\lambda$6584 and [SII]~$\lambda\lambda$\,6717,6731 
in the high-resolution spectrum, indicating 
an outburst of the star. Though several features observed in the G\,1200\,R
spectrum are characteristic of classical T~Tauri stars,
the nature of this star remains somewhat uncertain, because
it has a position in the {\sl H$-$K\/} vs. {\sl J$-$H\/} 
colour-colour diagram (see Fig.~\ref{Fig4})  where PMS stars 
are unexpected to be found.  Unfortunately no photometric data other than 
the 2MASS is available for this star, although data on photometric 
variations, accompanying the spectroscopic variation, might be helpful in
clarifying the nature of the star and its outburst, as well as could explain 
its position in the two-colour diagram.

The lithium absorption line is clearly absent from the spectra of 
2MASS J\,05060301$-$0715472 and J\,05060913$-$0712394. The only 
emission line in the spectra of these M-type stars is H$\alpha$, 
indicating that, though they are projected onto the molecular clouds,
they are not PMS stars.

2MASS~J\,05060574$-$0646151 and  05094864$-$0906065 are probably 
weak-line T~Tauri stars, though their nature have to be confirmed with
more reliable lithium observations. 2MASS~J\,05060574$-$0646151 is projected 
on the molecular cloud  G\,206.4$-$26.0, its H$\beta$ and  H$\alpha$ 
lines are filled with emission and the LiI\,$\lambda$\,6707 feature can be 
seen in its high-resolution spectrum.
The low resolution spectrum of 2MASS~J\,05094864$-$0906065, lying 
well outside the IC\,2118 molecular clouds, displays weak emission 
in H$\alpha$ and H$\beta$, and the $\lambda$\,6707 absorption can be
recognized in the high-resolution spectrum. 

\subsubsection{ALFOSC spectroscopy}
\label{Sect_ALFOSC}

Observations with ALFOSC revealed seven H$\alpha$ emission objects 
closely confined to the reflection nebula/molecular clouds. Four of
them are common with those observed with FLAIR. Two of the seven,
2MASS~J\,05060301$-$0715472 and 05060913$-$0712394 show neither lithium 
absorption, nor emission line other than H$\alpha$. These high S/N
observations confirm that they are field stars not related to
the clouds on which they are projected. The remaining five stars show 
both lithium absorption and 
emission lines characteristic of classical T~Tauri stars. 
Spectra of these objects are shown in Fig.~\ref{Fig2}.
The wavelength range of ALFOSC spectra was suitable for determining 
several flux ratios defined by Kirkpatrick et al.~(\cite{Kirk}) ({\sl A, B, C,
B\,/\,A, B\,/\,C\/}), Mart\'{\i}n \& Kun~(\cite{MK}) 
({\sl I$_2$, I$_3$\/}), and Preibisch, Guenther 
\& Zinnecker (\cite{PGZ}) ({\sl T1, T2\/}). We calibrated these 
spectral features against the spectral type and luminosity class
by measuring them in a 
series of standard stars observed during the same run. 
The accuracy of the two-dimensional spectral classification,
estimated from the range of spectral types obtained from different
flux ratios, is $\pm1$ subclass, except 2MASS J\,05071157$-$0615098,
whose spectrum shows extremely strong emission lines. The flux ratios
I$_2$ and I$_3$ are affected by the wing of the H$\alpha$ emission
line, whereas the wavelength range 7061--7088\,\AA, involved in $T_1$, 
contains the HeI emission line at 7065\,\AA, therefore these ratios
should not be used. The strong emission spectrum indicates high
accretion rate, therefore the effect of the veiling  has to 
be taken into consideration during the spectral classification. 
The hot continuum excess emission, originating from accretion shocks
on the stellar surface, makes the photospheric
absorption features shallower than they should be at the given 
effective temperature. All the spectral features applied during 
our classification increase in strength with decreasing effective 
temperature, therefore the effect of the veiling will be an apparently 
earlier spectral type. Its effect on the measured value of a flux ratio is 
difficult to quantify, because the broad wavelength intervals compared in 
the flux ratios contain several faint absorption features 
beyond the actually measured lines or bands, each affected by 
the veiling. Veiling is described as a
T=10000\,K blackbody, therefore its contribution is decreasing
with increasing wavelength. The spectral type obtained from 
the flux ratio {\sl A\/}, measuring the strength of CaH $\lambda$\,6975
feature may be more heavily influenced by veiling than that obtained 
from the NaI\,$\lambda\lambda$ 8183,8195 lines (flux ratio {\sl C\/}). 
We found no systematic difference between the spectral types derived 
from {\sl A, B, C, B\,/\,A, B\,/\,C\/}. The range of the derived spectral types
resulted in the accuracy of $\pm2$ subclasses for 2MASS J\,05071157$-$0615098.

Results of spectroscopy of PMS stars performed with ALFOSC are presented in 
Table~\ref{Tab3}. In addition to the derived spectral types we 
present the equivalent widths of the H$\alpha$ and LiI lines,
as well as indicate the additional emission lines observed in the
spectra. The uncertainties given in parentheses have been derived 
from the repeatability of the measurements. The real uncertainties of the 
LiI equivalent widths may be higher due to the blending of the line with 
neighbouring absorption or emission features (CaI\,$\lambda$6718, 
[SII]\,$\lambda$6717).
{\sl J\/} magnitudes, $J-H$ and $H-K_s$ colour indices 
from the 2MASS All Sky Catalog (IPAC~\cite{2MASS}) are also shown.
Both Fig.~\ref{Fig2} and Table~\ref{Tab3} clearly show that all these 
stars are cTTS.

\begin{figure*}
\centering{
\includegraphics[width=16cm]{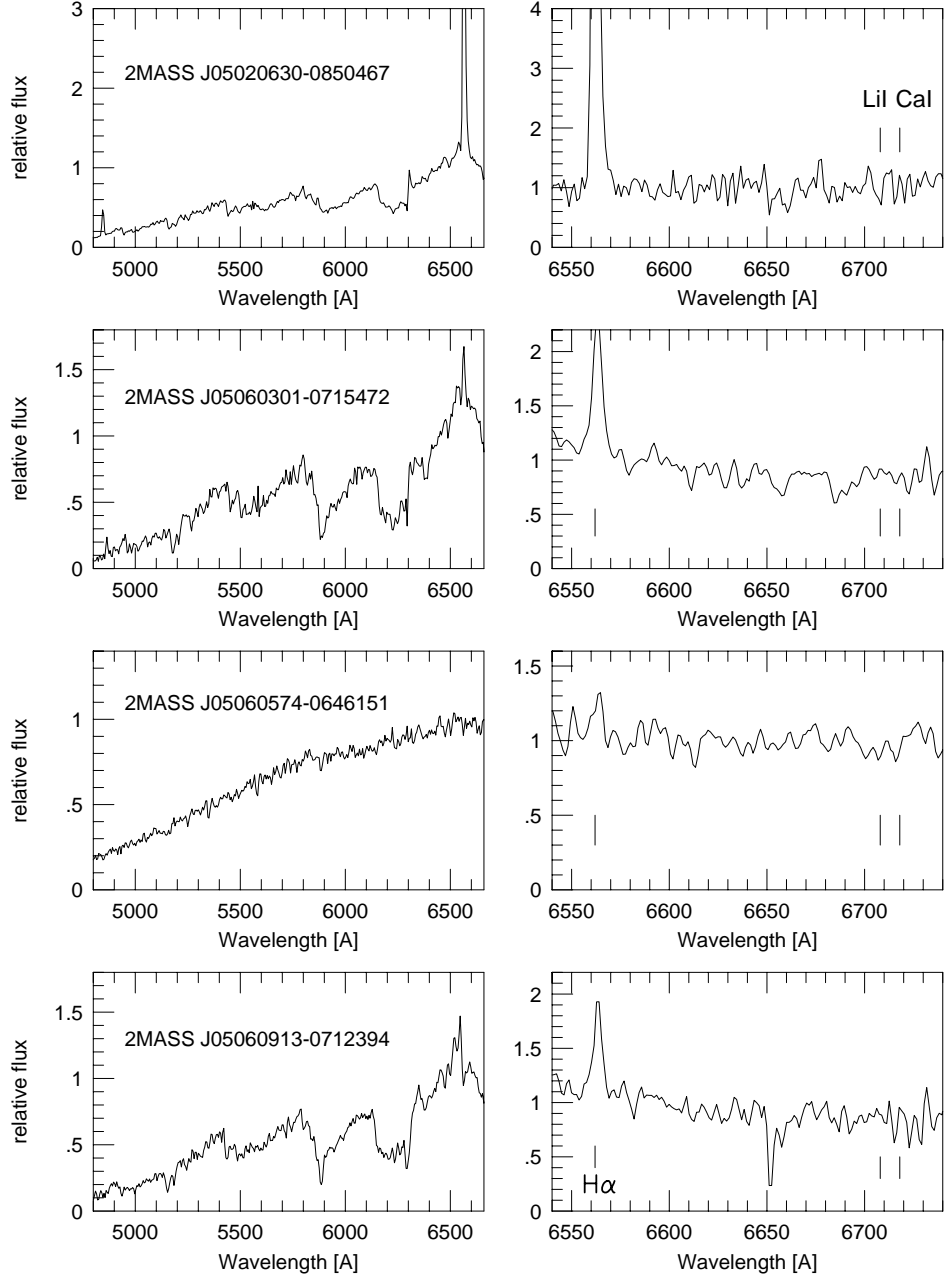}}
\caption{FLAIR spectra of H$\alpha$ emission objects found in the 
region centred on 5$^{\mathrm h}$0$^{\mathrm m}$, $-8\degr0\arcmin$. Left: 
low resolution spectra in the wavelength region 4800--6600\,\AA; right:
high resolution spectra in the region 6540--6740\,\AA. Positions of the 
H$\alpha$, Li\,I\,$\lambda$6707 and CaI\,$\lambda$6718  lines are indicated. 
No lithium absorption can be seen in
the spectra of 05060301$-$0715472 and 05060913$-$0712394.}
\label{Fig1}
\end{figure*}

\begin{figure*}
\setcounter{figure}{0}
\centering{
\includegraphics[width=16cm]{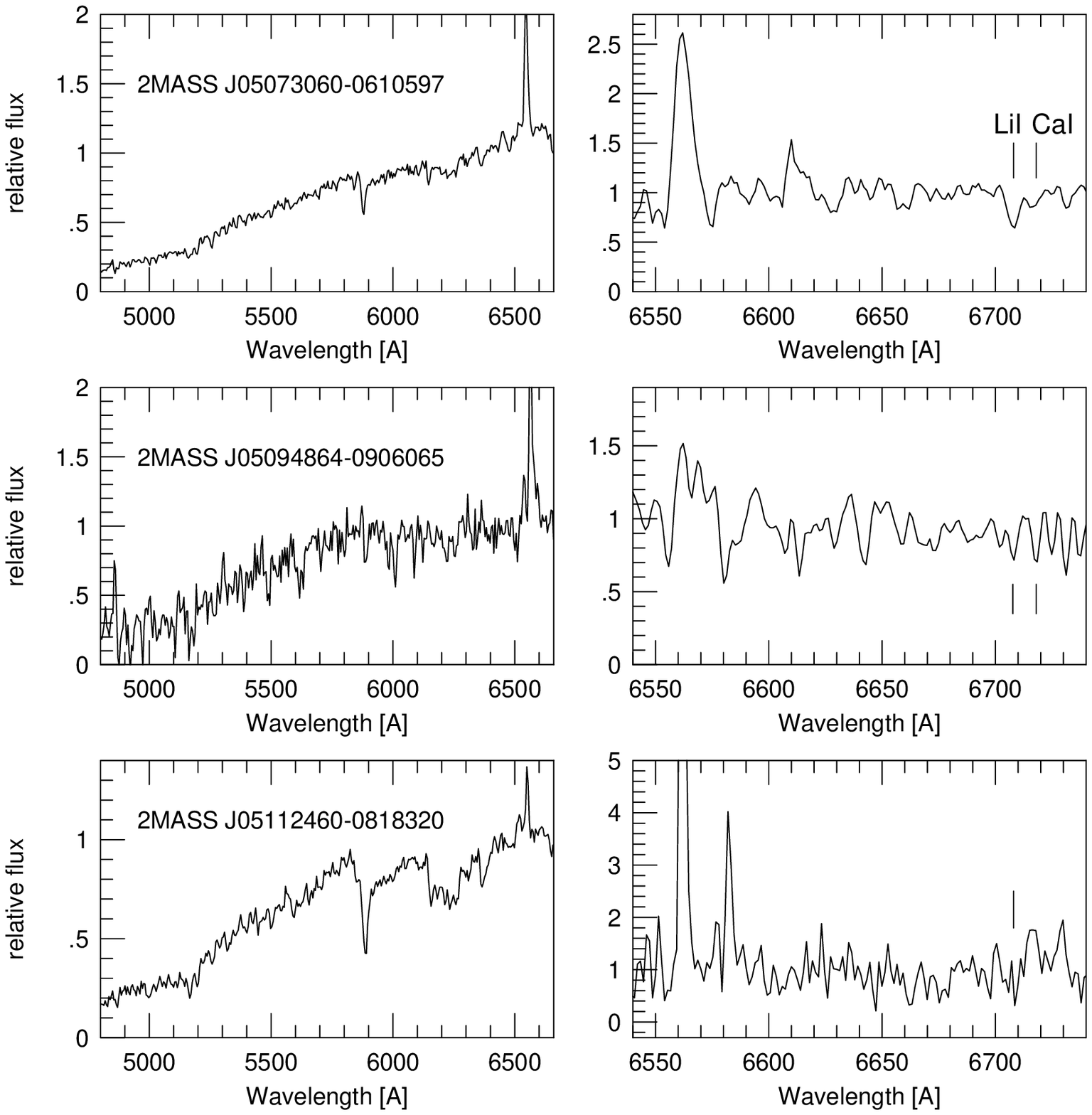}}
\caption{{\sl (Continued)}}
\label{Fig1}
\end{figure*}

\begin{figure*}
\centering{
\includegraphics[width=14cm]{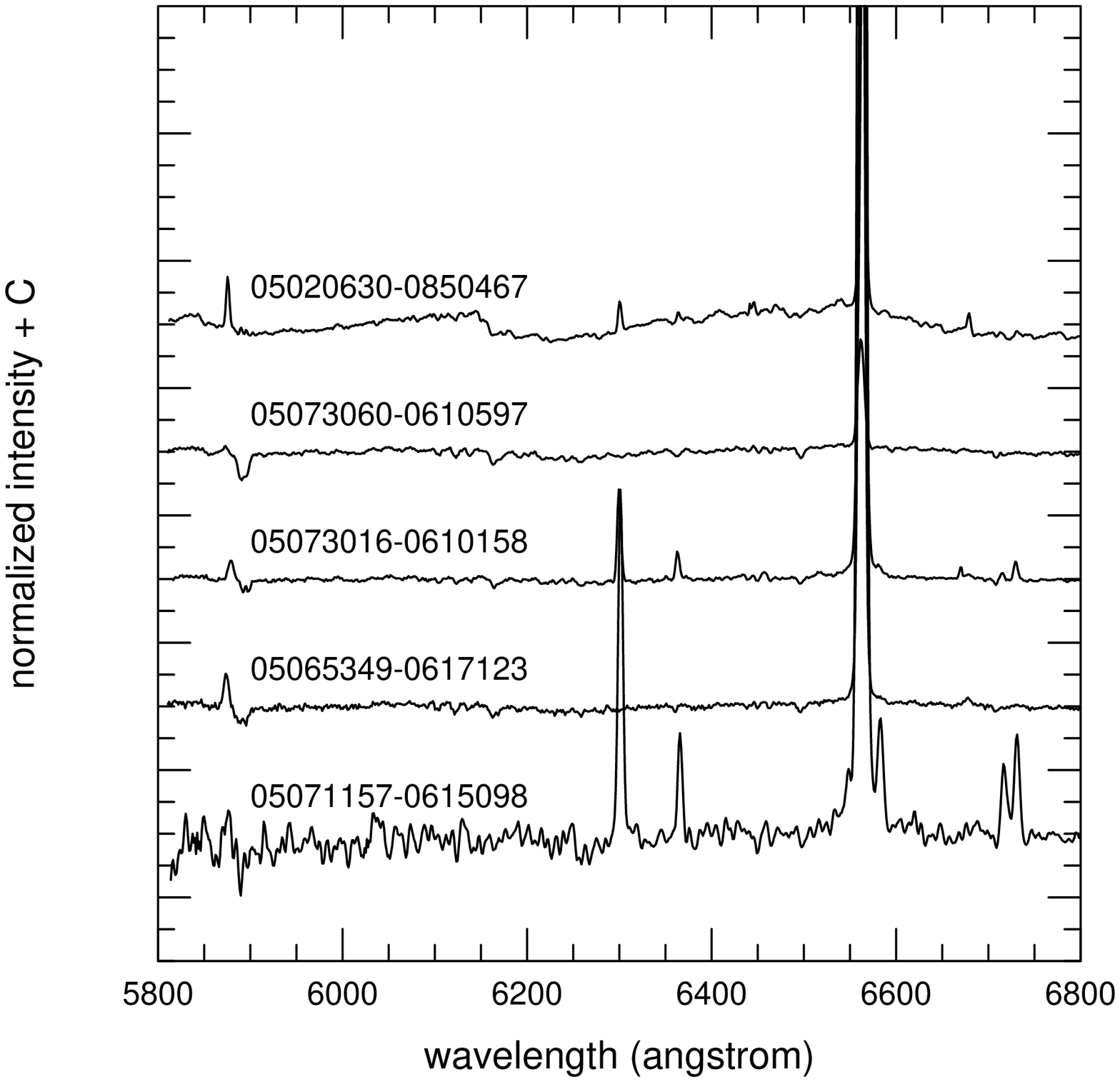}}
\caption{ALFOSC spectra of T~Tauri stars projected on the
clouds associated with IC\,2118.}
\label{Fig2}
\end{figure*}

\begin{table*}
\caption{Results of the spectroscopy and 2MASS data of the stars observed with FLAIR }
\label{Tab2}
\begin{flushleft}
\begin{tabular}{cl@{\hskip2mm}c@{\hskip2mm}c@{\hskip2mm}c@{\hskip1mm}r@{\hskip3mm}
c@{\hskip3mm}l}
\hline
\hline
2MASS J  & Sp.T. & $J$ & $J-H$ & $H-K_s$ & $W$(H$\alpha$) & LiI & Other emission lines \\
\noalign{\smallskip}
\hline
\noalign{\smallskip}
05020630$-$0850467 & M0 & 10.897 & 0.782 & 0.441 & $-$62.0\,(5.0) & yes & HeI, H$\beta$, [OI] \\ 
05060301$-$0715472 & M4 & 11.535 & 0.538 & 0.285 & $-$4.1\,(0.6) & no \\
05060574$-$0646151 & G8\,: & 11.749 & 0.700 & 0.272 & $-$1.87\,(0.6) & yes  \\
05060913$-$0712394 & M3 & 10.888 & 0.557 & 0.246 & $-$0.97\,(0.3) & no \\
05073060$-$0610597 & K7 & 10.127 & 1.081 & 0.743 & $-$13.4\,(1.0) & yes  \\ 
05094864$-$0906065 & G8 & 12.823 & 0.483 & 0.101 & $-$4.7\,(1.0)  & yes & H$\beta$ \\
05112460$-$0818320 & M0 & 12.159 & 0.756 & 0.144 & $-$30.0\,(2.0) & yes & S[II], [NI] \\
\noalign{\smallskip}
\hline
\end{tabular}
\end{flushleft}
\end{table*}

\begin{table*}
\caption{Pre-main sequence stars associated with IC\,2118: Results of ALFOSC spectroscopy 
and 2MASS data}
\label{Tab3}
\begin{flushleft}
\begin{tabular}{c@{\hskip2mm}l@{\hskip1mm}c@{\hskip1mm}c@{\hskip1mm}c@{\hskip1mm}r
@{\hskip3mm}c@{\hskip2mm}l}
\hline
\hline
\noalign{\smallskip}
2MASS J  & Sp.T. & $J$ & $J-H$ & $H-K_s$ & $W$(H$\alpha$) & $W$(LiI) & Other emission lines \\[3pt]
\noalign{\smallskip}
\hline
\noalign{\smallskip}
05020630$-$0850467 & M2IV & 10.897 & 0.782 & 0.441 & $-$58.4\,(0.8)\, & 0.29\,(0.03) & HeI, NaI \\ 
05065349$-$0617123 & K7IV & 11.182 & 1.255 & 0.754 & $-$112.4\,(5.0) & 0.45\,(0.03) & HeI \\
05071157$-$0615098 & M2IV  & 13.017 & 1.730 & 1.225 & $-$270.0\,(12.0) & 1.05\,(0.05) & [OI], [SII], HeI \\ 
05073016$-$0610158 & K6IV & 10.839 & 1.254 & 0.963 & $-$79.4\,(1.5) & 0.37\,(0.05) & [OI],
[NI], [SII], HeI, [FeII] \\ 
05073060$-$0610597 & K7IV & 10.127 & 1.081 & 0.743 & $-$13.5\,(1.0) & 0.46\,(0.04) & HeI \\ 
\noalign{\smallskip}
\hline
\end{tabular}
\end{flushleft}
\end{table*}

\begin{figure}
\resizebox{\hsize}{!}{\includegraphics{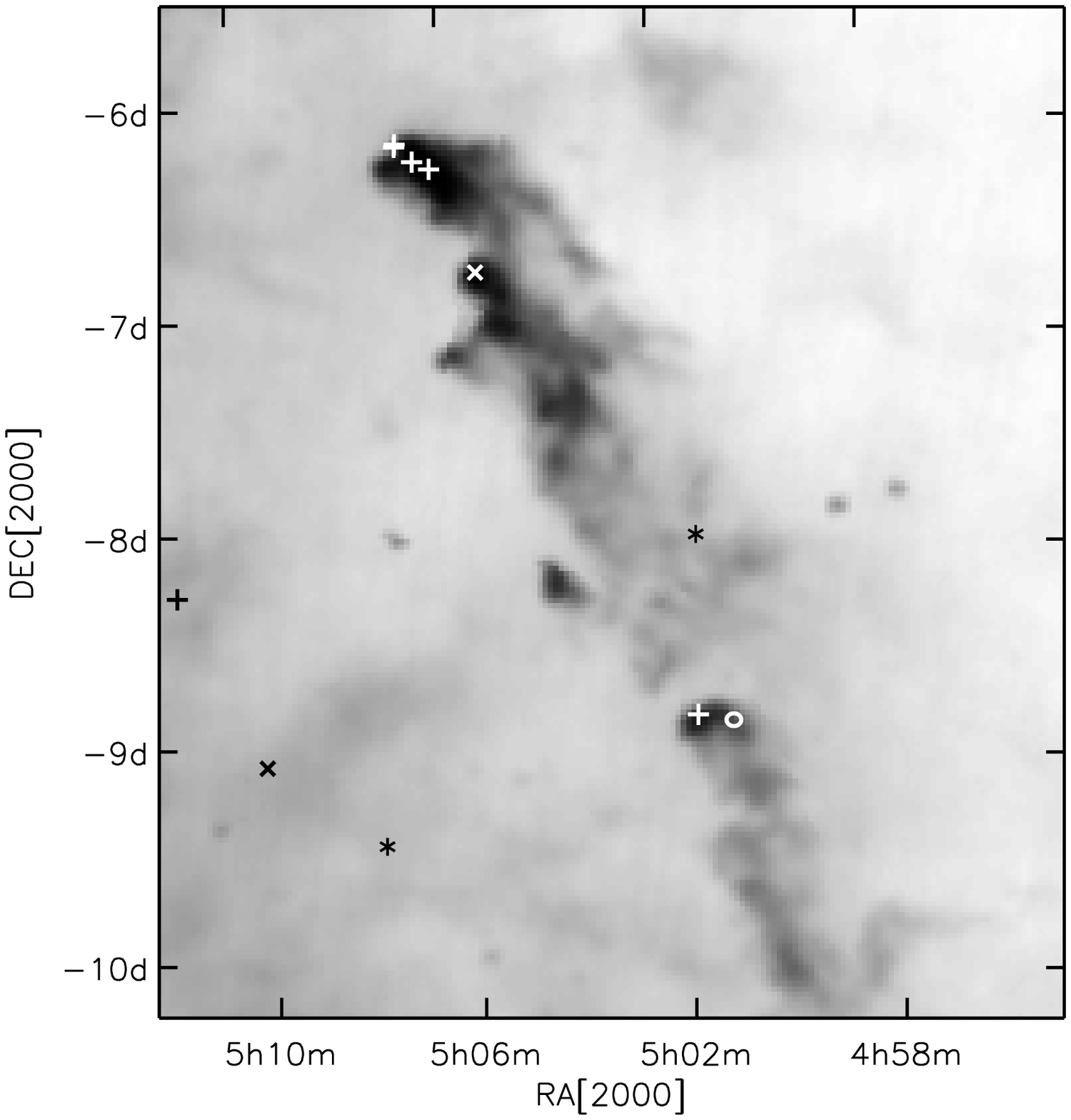}}
\caption{Distribution of the new PMS stars and other young 
stellar objects overlaid on the 100$\mu$m IRAS image of the region. 
`+' signs mark the new classical T Tauri stars (including the {\em candidate\/} cTTS
2MASS J\,05112460$-$0818320), and the candidate wTTS found during the 
present survey are marked with `x'. The open circle 
shows the position of IRAS~04591$-$0856 (HHL~17). Asterisks mark the weak-line
T Tauri stars identified by Alcal\`a et al.~(\cite{A96}) in the ROSAT all-sky 
survey data base.}
\label{Fig3}
\end{figure}

\section{Discussion}
\label{Sect_3}

Figure~\ref{Fig3} shows the surface distribution 
of the newly found PMS stars, together with other known 
YSOs of the region, overlaid on the IRAS 100$\mu$m image. In addition to 
the objects listed in Tables~2 and 3 two wTTS, identified by Alcal\`a et
al.~(\cite{A96}), and the embedded YSO IRAS~04591$-$0856 are plotted. 
The five cTTS listed  in Table~3, IRAS~04591$-$0856, as well as the 
candidate wTTS 2MASS J\,05060574$-$0646151 are projected 
on the molecular clouds associated with IC\,2118, while RXJ\,0502.4$-$0744 
is projected against a lower density part of the cloudy region. 

\subsection{HRD of the target stars}
\label{Sec_3.1}

\begin{figure}
\resizebox{\hsize}{!}{\includegraphics{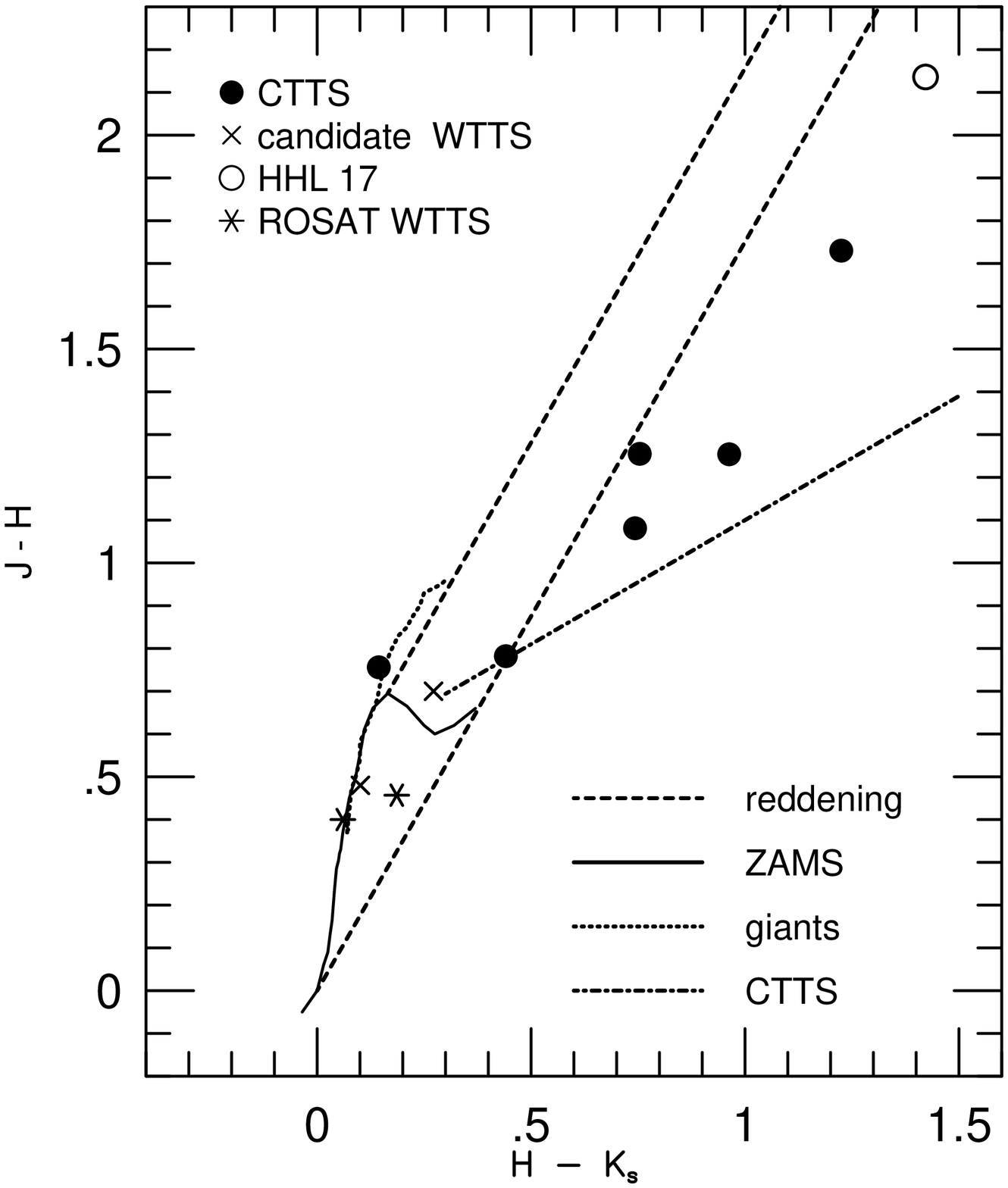}}
\caption[]{Positions of the PMS stars in IC\,2118 in the
$J-H$ vs. $H-K_s$ diagram. Loci of zero-age main sequence, giant branch,
and classical T\,Tauri stars, as well as the slope of interstellar
reddening are indicated. 2MASS\,J05112460$-$0818320, displaying cTTS-like
spectrum in the G\,1200R image, is located on the giant branch.}
\label{Fig4}
\end{figure}

We used $J$, $H$, and $K_s$ magnitudes 
obtained from the 2MASS All Sky Catalog~(IPAC~\cite{2MASS})
to place our stars on the Hertzsprung--Russell diagram.
For this purpose their effective temperatures and
bolometric luminosities are to be determined.
$T_{\rm eff}$ comes from the spectral type 
(Kenyon \& Hartmann~\cite{KH}), whereas $L_{\rm bol}$
can be determined from the near-infrared photometric data. 
Figure~\ref{Fig4} displays their positions on the 
$H-K_s$ vs. $J-H$ colour-colour diagram together
with the lines indicating the position of zero-age main-sequence, the
giant branch, direction of the interstellar reddening and the locus of
classical T~Tauri stars determined by Meyer, Calvet \& Hillenbrand (\cite{Meyer}).
In addition to the stars found during the present survey, the ROSAT 
wTTS (Alcal\`a et al.~\cite{A96} are also plotted.

The four cTTS associated with the cloud G\,206.4$-$26.0, and
HHL\,17 clearly display infrared excess, located to the right of 
the band of the reddened main sequence, whereas the positions of 
2MASS J\,05020630$-$0850467, associated with the cloud G\,208.4$-$28.3
and 2MASS J\,05060574$-$0646151, projected on G\,206.8$-$26.5, 
are equally compatible with unreddened cTTS and reddened 
main sequence stars. 
2MASS\,J05112460$-$0818320, which displayed the outburst during the 
FLAIR observing run, is also marked as cTTS in Fig.~\ref{Fig4}, 
though it lies on the giant sequence, rendering its nature somewhat uncertain. 

We made the widely used assumption that the total emission of
our target stars in the {\sl J\/} band originates from the
photosphere ({\sl e.g.}~Hartigan, Strom \& Strom~\cite{Hartigan}). 
Thus the colour index $J-H$ can be written as
$$ J-H = (J-H)_{0} + E_{\rm CS}(J-H) + E_{\rm IS}(J-H), $$
where $(J-H)_{0}$ is the true photospheric colour of the star,
$E_{\rm CS}(J-H)$ is the colour excess due to the emission from the
circumstellar disk in the {\sl H\/} band, and  $E_{\rm IS}(J-H)$ is
the colour excess originating from the difference of interstellar 
extinctions in the {\sl J\/} and {\sl H\/} bands.

We dereddened our cTTS onto the locus of unreddened T~Tauri stars in 
the $H-K_s$ vs. $J-H$ colour-colour diagram (Meyer et al.~\cite{Meyer}) 
in order to determine $E_{\rm IS}(J-H)$. Bolometric 
luminosities were derived from the {\sl J} magnitudes and $E_{\rm IS}(J-H)$
colour excesses by using the interstellar extinction law
$A_{\rm J}=2.65\,\times E_{\rm IS}(J-H)$
(Rieke \& Lebofsky~\cite{RL}), and the bolometric corrections tabulated
by Hartigan et al.~(\cite{Hartigan}). 

\begin{figure}
\resizebox{\hsize}{!}{\includegraphics{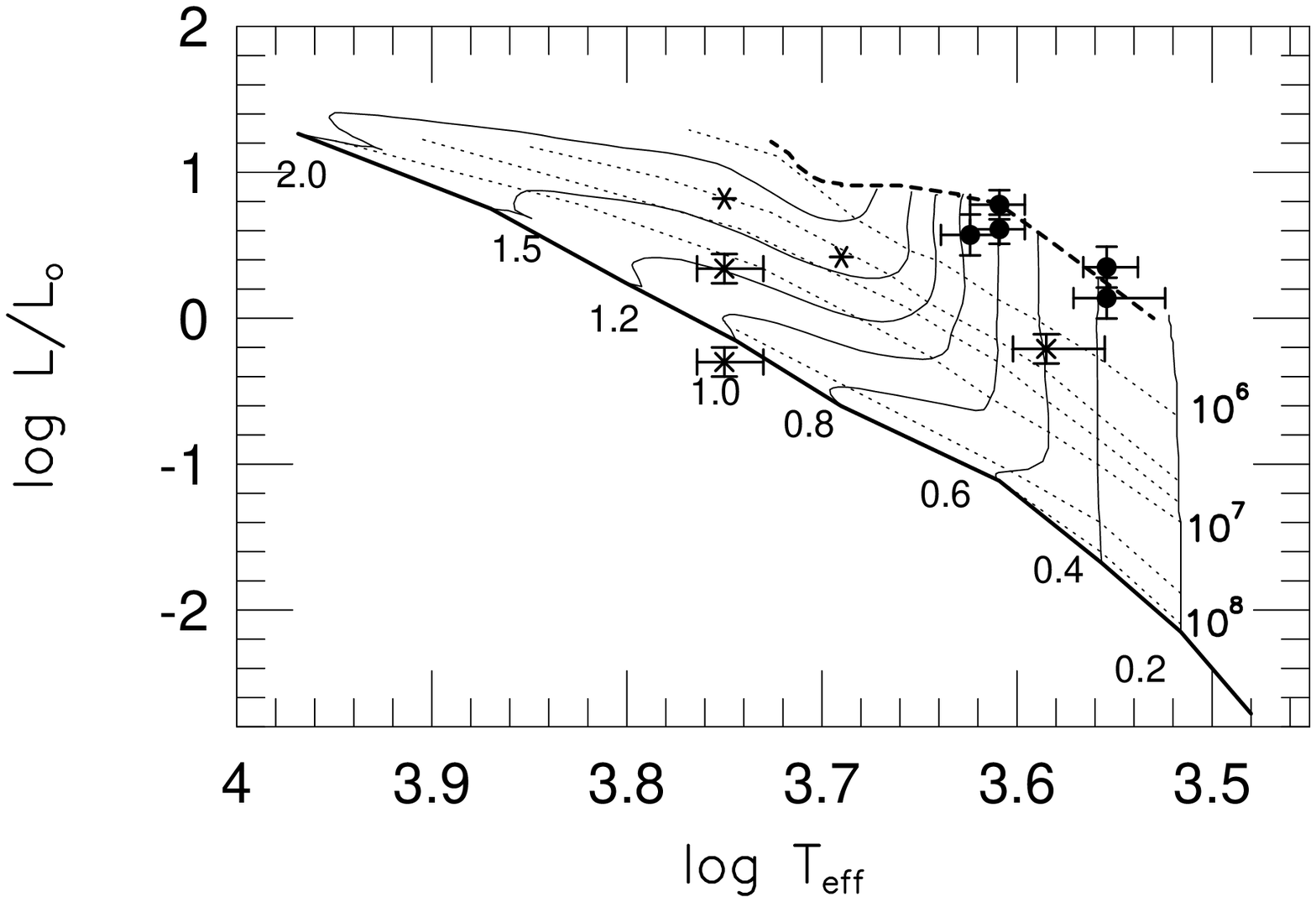}}
\caption{Positions of the observed stars in the HRD, assuming a distance
of 460\,pc. Black dots indicate classical T~Tauri stars associated with IC\,2118, 
crosses mark the  other target stars and  asterisks are for  
wTTS detected by ROSAT (Alcal\`a et al.~\cite{A96}). Dotted lines indicate 
the isochrones of 10$^6$, 3$\times10^6$, 5$\times10^6$ , 10$^7$, 5$\times10^7$ and 
10$^8$ years, and thin solid lines show the evolutionary
tracks from Palla \& Stahler's~(\cite{PS}) model. The dashed line
corresponds to the birthline and thick solid line indicates the
zero age main sequence.}
\vskip -0.4cm
\label{Fig5}
\end{figure}

\begin{figure}
\resizebox{\hsize}{!}{\includegraphics{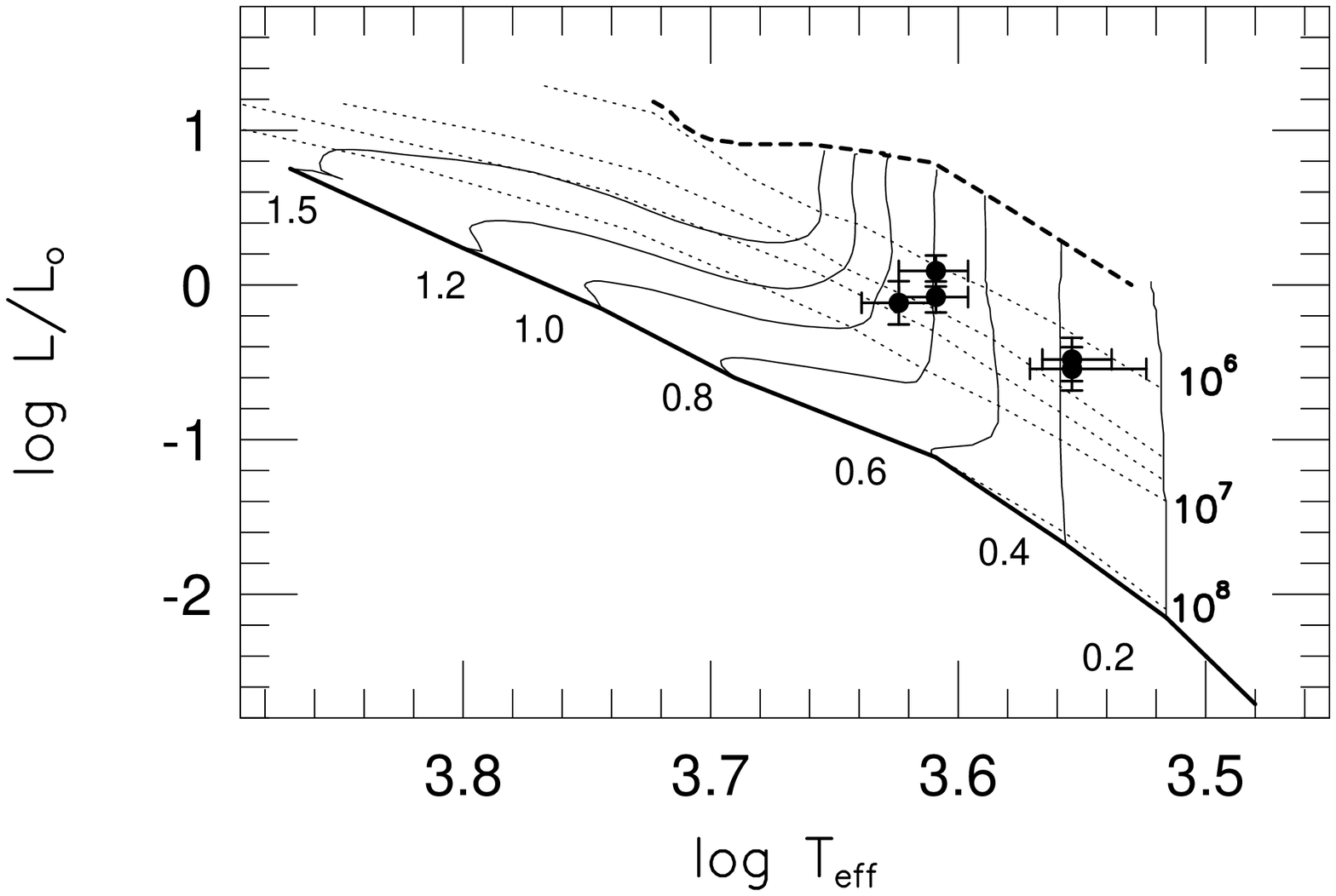}}
\caption{The T~Tauri stars of IC\,2118 in the HRD, assuming a distance
of 210\,pc. Isochrones and evolutionary tracks, as well as the birthline and
zero age main sequence are indicated as in Fig.~5.}
\vskip -0.4cm
\label{Fig6}
\end{figure}

The distribution  of our target stars in the HRD is displayed in Fig.~\ref{Fig5}, 
with the assumption that all of them are located at 460\,pc, 
at the distance of main Orion molecular clouds. Uncertainties of 
$\log\,T_\mathrm{eff}$  are derived from those of the 
spectral classification. In assessing the
uncertainty of $\log\,L$ the errors of photometric 
data, given in the 2MASS All Sky Catalog, and uncertainties of the
bolometric corrections due to the error of spectral classification
were taken into account. 
Further sources of uncertainty of $\log\,L$ are negligence of 
the excess luminosity arising from photospheric veiling and circumstellar 
dust emission. Both of these effects, however, have their minima around 
1\,$\mu$m (Kenyon \& Hartmann~\cite{KH}).
Evolutionary tracks and isochrones, as
well as the position of the birthline and zero-age main sequence
(Palla \& Stahler~\cite{PS}) are also shown. However, distances of the
stars, in particular of those outside the IC\,2118 molecular clouds, are actually  unknown. 
They may  either be low-mass members of Orion~OB1, closer to us than the A and
B clouds, or may be situated 
at different distances as members of the Gould Belt system 
(Alcal\`a et al.~\cite{ACT}). Therefore this figure only indicates 
that if they are as distant as the giant clouds of most recent star 
formation, then most of them are PMS stars at different evolutionary 
stages. The only exception is 2MASS J\,05094864$-$0906065, located far 
from the IC\,2118 clouds on the sky, and below the ZAMS in Fig.~\ref{Fig5}.
This star is probably more distant ($d \ga 800$pc) than the Orion star 
forming region, given that H$\alpha$  and H$\beta$ emission, seen in
its spectrum, are indicative of PMS nature for this spectral type.
In this figure the five cTTS found in the IC\,2118 molecular 
clouds are located high above the 1\,Myr isochrone, around 
the birthline. It was shown by Baraffe et al.~(\cite{BCAH02}) 
that stellar ages and evolutionary tracks are very uncertain 
at this part of the HRD. The birthline shown here is considered  
as an upper limit for pre-main sequence luminosities, 
and even it is probable that the youngest accreting low-mass stars 
appear below this line (Hartmann et al.~\cite{HCK}). 
Four of the five cTTS are located at the outer regions of the dense 
C$^{18}$O cores of their parent clouds (Yonekura et al.~\cite{Yonekura}), 
suggesting that they  have already evolved off the birthline. 
Therefore their positions  in this diagram provide  further
support for the result that IC\,2118 is closer to us than Orion A and B.

Adopting that IC\,2118 is located at a distance 
210\,pc from us, the HRD shown in Fig.~\ref{Fig6} has been obtained. 
In this plot the cTTS projected on IC\,2118 form a group in the mass 
interval 0.4--0.9\,M$_{\odot}$, and they  are scattered between 
the isochrones of 1$\times10^{6}$  and 4$\times10^{6}$ years.
The candidate wTTS seen along the line of sight of a molecular cloud, 
2MASS J\,0506057$-$0646151, is probably 
a more distant object, lying  behind the clouds.
Properties of these five T~Tauri stars, resulted from our study, are shown in 
Table~\ref{Tab4}. Effective temperatures $T_\mathrm{eff}$ corresponding to the spectral
classes  are displayed in column 2, and visual extinctions  
$A_{\rm V}=9.14\,\times E_{\rm IS}(J-H)$ (Rieke \& Lebofsky~\cite{RL}) 
are shown in col. 3. Luminosities  derived from the 2MASS data are listed in col. 4, and
masses and ages  resulting from the Palla \& Stahler~(\cite{PS}) 
model, are shown in columns 5 and 6, respectively. Minimum and maximum values
of the derived quantities, resulting from the errors of spectral classification and photometry,
are indicated in parentheses.

\begin{table*}
\caption{Properties of the pre-main sequence stars associated with IC\,2118, 
derived from spectroscopic and 2MASS data}
\label{Tab4}
\begin{flushleft}
\begin{tabular}{cccccc}
\hline
\hline
\noalign{\smallskip}
2MASS J  & $T_\mathrm{eff}$ & $A_V$ &  $L$ & $M$ & Age  \\
 &  (K) & (mag) & ($L_{\sun}$) &  ($M_{\sun}$) & (10${^6}$ yr) \\
\noalign{\smallskip}
\hline
\noalign{\smallskip}
05020630$-$0850467 & 3580\,($^{3720}_{3470}$) & 0.0\,($^{+0.5}_{-0.5}$) &  0.33\,($^{0.42}_{0.27}$)  &
0.38\,($^{0.45}_{0.30}$) & 1.5\,($^{2.0}_{1.0}$) \\ [.5ex]
05065349$-$0617123 & 4060\,($^{4205}_{3955}$) & 4.0\,($^{4.7}_{3.4}$) &  0.84\,($^{1.04}_{0.66}$)  &
0.80\,($^{0.90}_{0.66}$) & 2.5\,($^{3.0}_{2.0}$) \\[.5ex]
05071157$-$0615098 & 3580\,($^{3850}_{3370}$) & 6.8\,($^{7.8}_{6.3}$) & 0.29\,($^{0.36}_{0.23}$) &
0.38\,($^{0.45}_{0.22}$)  & 2.0\,($^{3.0}_{0.9}$) \\ [.5ex]
05073016$-$0610158 & 4205\,($^{4350}_{4060}$) & 2.5\,($^{3.1}_{2.1}$) & 0.77\,($^{0.93}_{0.63}$)  &
0.90\,($^{1.05}_{0.80}$) &  4.0\,($^{5.5}_{2.5}$) \\ [.5ex]
05073060$-$0610597 & 4060\,($^{4205}_{3955}$) & 1.7\,($^{2.4}_{1.2}$) & 1.23\,($^{1.56}_{1.00}$)  &
0.80\,($^{0.90}_{0.66}$) & 1.0\,($^{1.6}_{0.9}$)  \\ 
\noalign{\smallskip}
\hline
\end{tabular}
\end{flushleft}
\end{table*}

\subsection{The IC\,2118 association}
\label{Sect_3.2}

Both the surface distribution of the newly identified cTTS and their
position in the HRD suggest the presence of a young association of low-mass
stars formed in the high latitude molecular clouds associated with 
IC\,2118. The five cTTS identified in this work are projected on
two different molecular clouds. The cloud G\,206.4$-$26.0 hosts
four of the stars. The mass of this cloud is 85\,M$_{\sun}$ (Paper~I),
and its radial velocity of $v_\mathrm{LSR} \approx -2.2$\,kms$^{-1}$,
significantly more negative than those of Orion~OB1 and Orion~A and B,
suggesting that it represents a distinct subsystem of the Orion
star forming region.
It contains an elongated dense core mapped in C$^{18}$O 
by Yonekura et al.~(\cite{Yonekura}). The mass of the cores, traced by 
C$^{18}$O, is 25\,M$_{\sun}$ (scaled to 210\,pc Yonekura et al.'s~(\cite{Yonekura})
result). The stars associated with this
cloud are aligned parallel to the long axis of the core, at a mean projected
distance of $\sim\,0.3$\,pc from each other, so that the 
two nebulous objects in RNO\,37 as well as 05065349$-$0617123 are located at the 
outskirts, while the 05071157$-$0615098 is projected 
inside the core (see Fig.~\ref{Fig3}: the IRAS 100$\mu$m intensities
show largely the same structure  as $^{13}$CO and C$^{18}$O maps).
This surface distribution suggests an age sequence: the star closer to the centre
of the core should be younger. This age sequence  is washed out by 
the uncertainties of $T_\mathrm{eff}$ and $L_\mathrm{bol}$, but 
the  signposts of strong accretion, observed in the spectrum of 05071157$-$0615098
may be indicative of its extreme youth indeed. This star may be significantly
younger than the age derived from its position in the HRD.
Both theoretical and observational studies suggest that strongly accreting 
PMS stars may be considerably less luminous than their coeval, non-accreting
counterparts, mimicking an older age (Hartmann et al.~\cite{HCK}; 
Comer\'on et al. \cite{CFBNK}).

Comparison of the $J$ magnitudes of the 2MASS and
DENIS~(DENIS Consortium~\cite{DENIS}) data bases, moreover, reveals the 
variability of this star: contrary to the 2MASS magnitude $J=13.017\,\pm0.026$, 
the same value for DENIS J\,050711.5$-$061509 is
$J=12.693\,\pm0.08$\,mag. The variability in the $J$ band also 
contributes to the uncertainty of the derived luminosity. 
 
The fifth member of the IC\,2118 association, 2MASS 
J\,05020630$-$0850467 is projected on the molecular 
cloud G\,208.3$-$28.4 (MBM\,21), whose mass was estimated 
to be 14\,M$_{\sun}$  (Paper~I). The radial velocity of this cloud
is $v_\mathrm{LSR} = +4.8$\,kms$^{-1}$, close to the average
value of Orion~OB1. The large velocity difference between the 
northern and southern clouds of the IC\,2118 complex may suggest
that they are unrelated objects at different distances. This is unlikely
because both the illumination of the northern clouds by Rigel and
the distance determination for the southern cloud by Penprase~(\cite{Penprase})
converge to the same distance value adopted here. One may notice, however,
that the velocity pattern of the IC\,2118 complex is similar to that
observed by Bally~(\cite{Bally89}) in Orion~A:  
while the radial velocities of the southern parts of both Orion~A and 
IC\,2118 are nearly the same as that of Orion OB\,1, the northern
portions  have  more positive velocities in Orion~A
and  more negative velocities in IC\,2118. Both regions
are located to the south of Ori~OB1a, the centre of the expansion of the 
Orion--Eridanus Bubble; Orion~A resides in the receding hemisphere, and IC\,2118
in the approaching one. Thus the observed velocity structures suggest
that the northern parts of the clouds, closer to Ori~OB1, have experienced greater 
acceleration  than those farther from the origin of the
shock wave, compressing and subsequently accelerating the clouds.

The cloud contains two dense C$^{18}$O cores having masses of 7.7 and 
3.5\,$M_{\sun}$, respectively (scaled to 210\,pc the values 
derived by Yonekura et al.~\cite{Yonekura}). The star is located at the
edge of the larger, eastern core, whereas the smaller, western core
contains the embedded infrared source IRAS~04591$-$0856.

The large-scale geometry and kinematics of the Orion--Eridanus region
suggests that star formation in the IC\,2118 region propagates from
the north-east toward the south-west, and also toward us. According
to this picture, the two YSOs associated with  G\,208.3$-$28.4 are
probably somewhat younger than their counterparts in G\,206.4$-$26.0. 
Both the derived age of  J\,05020630$-$0850467  and the deeply embedded 
state of IRAS\,04591$-$0856 support this hypothesis.

G\,208.3$-$28.4 is one of the smallest known star forming molecular
clouds in our galactic neighbourhood. Its star forming cores contain significantly
less material than the average C$^{18}$O mass of $\approx$12\,M$_{\sun}$, required 
to form a protostar in Taurus and Chamaeleon  (Onishi et al.~\cite{Onishi};
Mizuno et al.~\cite{Mizuno}). This may be the consequence of the high ambient
pressure from the superbubble, leading to a smaller critical 
mass for gravitational collapse.

The velocity gradient along the cloud complex suggests that 
the compressed clouds are subsequently accelerated by the shock propagating
from Ori~OB1, and that their acceleration continues after the onset of star 
formation. The probability of observing YSOs in very small, compressed clouds is
low not only because small clouds disperse rapidly, but also because
they are swept off the newly formed stars, which keep their velocities
while their parental clouds are further accelerated.  

Alcal\`a et al.~(\cite{ACT2000}) identified a subsample of  wTTS
widely distributed over the Orion region with radial velocities 
$v_\mathrm{LSR} < +6$\,kms$^{-1}$,
among them are the two stars, RXJ\,0502.4$-$0744 and  RXJ\,0507.8$-$0931, located
within the field studied here. The presence of the young stars 
in G\,208.3$-$28.4 gives some support to the speculation
that these stars might have been born in small clouds
compressed to form stars and then swept aside and dispersed by the approaching 
hemisphere of the superbubble. In this case the distance of these stars 
should be between 200--350\,pc. The positions in the HRD of both 
RXJ\,0502.4$-$0744 and  RXJ\,0507.8$-$0931 favour the
higher limit, because, according to the scenario of sequential star formation, 
they should be younger than $10^7$ years.

It was established in Paper~I
that the clouds associated with IC\,2118 lie on the surface of the 
Orion--Eridanus Bubble, being blown with variable powers by the  
stellar winds and supernova explosions of the massive stars 
of Orion~OB1 during the last ten million years. The ages of 
the PMS stars found in the clouds are compatible
with the assumption that star formation has been triggered by
the superbubble. The complicated geometry and wind history of the 
OB association (Brown et al.~\cite{BHB}) hinders both any detailed 
speculation on the exact position and age of the sources of 
trigger and any accurate mapping of the shape of the bubble surface. Wherever the 
shock wave meets a dense medium, a new section of surface will arise.  

The stars found during the present studies are probably the most
massive members of the young stellar group born in the low-mass,
high-latitude molecular clouds. Several faint and red 2MASS and 
DENIS sources are projected on the clouds, whose nature is uncertain 
due to the low S/N of the data. Further members of the
IC\,2118 association can be revealed by spectroscopic and deep 
near infrared observations of these sources.

\section{Summary of results}
\label{Sect_4}

We identified a new nearby association of classical T~Tauri stars
in the region of the reflection nebula IC\,2118, at a distance of
210\,pc from the Sun. Our spectral classification, together 
with near-infrared photometric data published in the 2MASS All Sky 
Catalog (IPAC~\cite{2MASS}), allowed us to derive effective 
temperatures and luminosities of these stars. 
Comparison of these data  with theoretical pre-main 
sequence evolutionary tracks has shown that masses of the association 
members are in the interval 0.4--0.9\,M$_{\sun}$, and their ages are 
in the interval 1--4\,Myr. Our results suggest that star formation
was triggered in the IC\,2118 clouds by shock waves originating from
Orion~OB\,1, and thus this group of young low-mass stars is a 
distinct subsystem of the Orion star forming region.

\begin{acknowledgements}

We are indebted to Quentin Parker and Paul Cass for their help in obtaining
FLAIR data, and to Francesco Palla for sending his data set on pre-main 
sequence evolution.  
This work is partly based on observations with Nordic Optical Telescope  
operated on the island of La Palma jointly by Denmark, Finland, 
Iceland, Norway, and Sweden, in the Spanish Observatorio 
del Roque de los Muchachos of the Instituto de Astrofisica de Canarias.
The data presented here have been taken using ALFOSC, which is owned
by the Instituto de Astrofisica de Andalucia (IAA) and operated at the
Nordic Optical Telescope under agreement between IAA and the NBIfAFG of 
the Astronomical Observatory of Copenhagen.
This publication makes use of data products from the Two Micron All Sky Survey, 
which is a joint project of the University of Massachusetts and the Infrared 
Processing and Analysis Center/California Institute of Technology, funded by 
the National Aeronautics and Space Administration and the National Science Foundation.
We also utilized DENIS data DENIS has been supported financially mainly by the 
French Institut National des Sciences de l'Univers, CNRS, and French 
Education Ministry, the European Southern Observatory, 
the State of Baden-Wuerttemberg, and the European Commission under 
networks of the SCIENCE and Human Capital and Mobility programs, the 
Landessternwarte, Heidelberg, l'Institut d'Astrophysique de Paris, 
the Institut f\"ur Astrophysik der Universit\"at Innsbruck and Instituto 
 de Astrofisica de Canarias. \\
Financial support from the Hungarian OTKA grants T34584 and T37508,
and from the Serbian P1191 grant is acknowledged. The paper is benefited
from the comments of the anonymous referee.
\end{acknowledgements}

\end{document}